\documentclass[twocolumn,showpacs,preprintnumbers,amsmath,amssymb]{revtex4}
 \usepackage{graphicx}
 \usepackage{dcolumn}
 \usepackage{bm}
\begin{document}


\title{Comment on "Two Distinct Quasifission Modes in the
$^{32}$S+$^{232}$Th Reaction" \\(by D.J.Hinde et al. PRL 101, 092701 (2008))}
\author{G.G.Adamian$^{1,2}$, N.V.Antonenko$^{1}$}
\affiliation{$^{1}$Joint Institute for Nuclear Research, 141980 Dubna, Russia\\
$^{2}$Institute of Nuclear Physics, 702132 Tashkent, Uzbekistan
}


\maketitle

In a recent Letter \cite{Hinde}, D.J.Hinde {\it et al.} presented comprehensive
measurements of fission-type cross sections, angular anisotropies, mass distributions, and
mass-angle distributions for the $^{32}$S+$^{232}$Th reaction at various bombarding
energies. One of the interesting conclusions of this Letter is that the quasifission (QF)
dominates in the formation of more mass-symmetric component in the mass distribution.
This conclusion partly arises from the comparison
with the prediction of the transition state model.
Although the decisive role of QF for the mass-symmetric
part of the mass distribution  was theoretically predicted in Refs.~\cite{Quaal,Yapon},
the experimental arguments, in our opinion, are still   insufficient.

Similar mass-angle distributions were measured already long time ago
and the correlation of the fragment mass with
angle was seen as well, for example in Refs.~\cite{Toke0}.
In Refs.~\cite{Toke0} the triple-differential cross sections,
$d^3\sigma/dAd\theta_{c.m.}dTKE$, were obtained for the binary events within the full
range of mass $A$ and total kinetic energies (TKE) and within almost the full range of
the center-of-mass angle $\theta_{c.m.}$. In Refs.~\cite{Toke0} an observed distinct dependence
between fragment mass and scattering angle shows that new reaction
channel comes into play, namely the quasifission process. The central feature of
this reaction mechanism is the evolution of the reaction complex towards mass symmetry.
The experiments of Refs.~~\cite{Toke0} in 1984-1987 seem to be more informative than the
experiment in Ref.~\cite{Hinde}, because in Ref.~\cite{Hinde}
the TKE of fragments is not presented and the experimental uncertainties of the separation
of quasifission events from quasi-elastic and deep inelastic
events for the mass asymmetric component are not discussed.

The systematical experimental study of the reactions
with various  charge asymmetries in the entrance  channel
but with the same total charge number is required to conclude reliably
on the role of QF in the formation of mass-symmetric part of the
mass (charge) distribution. The role of QF seems to be strongly increased with decreasing charge asymmetry  in the
entrance channel of the reaction \cite{Nashi,ikh}.
One can compare the mass (charge) and mass-angle distributions
measured in  reactions
$^{16}$O+$^{249}$Cf, $^{20}$Ne+$^{245}$Cm,
$^{32}$S+$^{232}$Th, $^{40}$Ar+$^{226}$Ra, $^{54}$Cr+$^{208}$Pb,
$^{64}$Ni+$^{198}$Pt, $^{70}$Zn+$^{192}$Os, and $^{76}$Ge+$^{186}$W
which all lead to  almost the same compound nuclei.
The reaction combinations with large and small
charge asymmetries would
have different characteristics of fission-type products.
Comparing the yields of near-symmetric mass splits and the evaporation residue cross
sections in the reactions, for example, $^{54}$Cr+$^{208}$Pb and $^{64}$Ni+$^{198}$Pt,
one can unambiguously separate the fusion-fission and QF processes.
%
The comparison of the small yields of the products
with  $Z=6-16$  in the $^{54}$Cr+$^{208}$Pb and $^{64}$Ni+$^{198}$Pt reactions can give  the
complementary information \cite{Quaal} about the QF process.

The authors of Letter \cite{Hinde} wrote that "With the present lack of a realistic dynamical
model of QF with arbitrary oriented deformed fragments, only qualitative explanations
for the two distinct QF components can be proposed". In our opinion this statement
is not correct and misleads readers. In Refs.\cite{Quaal,Quaal1,Yapon} the dynamical models
were suggested to describe the properties of quasifission. As follows from Ref.\cite{Quaal} (see Table~I)
and from Ref.\cite{Yapon}, the QF mainly contributes to the formation of
mass-symmetric component
that is in agreement with  the conclusions (iv)--(vi) of Letter \cite{Hinde}.

As follows from the conclusion (iii) \cite{Hinde}, the compact antialigned
contact configuration, which is formed with larger probability at energies larger than the value of the Coulomb barrier,
is more favorable for fusion and because of this the probability  of asymmetric QF component decreases.
However, the Fig.~3 of Letter \cite{Hinde} and the conclusions (iv)--(vi)
result that the probability of QF symmetric component grows with increasing bombarding energy.
It is clear that the increase of the relative contribution of the fusion-fission
with respect to the QF should influence
by the same way on symmetric and asymmetric  components.
So, the conclusion (iii) is not well justified.

In Ref.~\cite{Quaal} the experimental energy distribution of quasifission fragments
were analysed to find the most probable orientation of the nuclei in the dinuclear
system. The orientation of the nuclei seems to be important for the capture stage
and the formation of initial dinuclear system. However, the maximum of energy distribution
of mass-symmetric component weakly depends on the bombarding energy and, thus, on the
orientation of nuclei in the entrance channel.
The authors of  Ref.~\cite{Hinde} discussed much the orientation effect in the entrance
channel but they did not present the TKE of fragments.

In our opinion the  maximum in the asymmetric mass  yields arises
from the minima on the potential energy surface
and is caused by the shell effects around nuclei $^{208}$Pb and $^{66}$Ni.
For low excitation energy,  the evolution of the dinuclear system towards symmetry is hindered
by these minima.
With increasing excitation or bombarding energy the mass flow towards symmetry
increases and the fractional yield of mass-asymmetric component falls
but the fractional yield of mass-symmetric component growths. Since
at the same time the fusion probability
grows, we have only the redistribution of
mass-asymmetric and -symmetric QF components vanishing
the asymmetry of the mass distribution.
Comparing the experimental mass distributions \cite{Toke0,Toke} at different bombarding energies,
one can see that the mass drift towards symmetry increases with $E_{{\rm c.m.}}$
smearing the asymmetric QF component.
In order to separate the effects of deformation in the entrance channel from
the shell effects of the potential energy surface, one can compare the QF products
of the reactions having  different (small and large) entrance channel deformation and
 forming the same compound nucleus in the complete fusion process:
$^{40,48}$Ca+$^{152,144}$Sm  or
$^{80,86}$Kr+$^{150,144}$Sm.



\begin{thebibliography}{99}
\bibitem{Hinde} D.J.Hinde, R.du Rietz, M.Dasgupta, R.G.Thomas,
L.R.Gasques, Phys.Rev.Lett. {\bf 101}, 092701 (2008).
\bibitem{Toke0} G.Guarino {\it et al.}, Nucl.Phys.A {\bf 424}, 157 (1984);
J.T\"oke {\it et al.}, Nucl.Phys.A {\bf 440}, 327 (1985);
W.Q.Shen {\it et al.}, Phys.Rev.C {\bf 36}, 115 (1987).
\bibitem{Quaal}
G.G.Adamian, N.V.Antonenko,  W.Scheid,
Phys.Rev.C {\bf 68}, 034601 (2003).
\bibitem{Yapon}
Y.Aritomo, M.Ohta,
Nucl. Phys. A {\bf 744}, 3 (2004).
\bibitem{Nashi}
 N.V.Antonenko {\it et al.},
Phys.Rev.C {\bf 51} (1995) 2635;
G.G.Adamian, N.V.Antonenko, W.Scheid,
V.V.Volkov, Nucl.Phys.A {\bf 627}, 361 (1997); {\bf 633}, 409 (1998);
G.G.Adamian, N.V.Antonenko,   W.Scheid,
Nucl.Phys.A {\bf 678}, 24 (2000).
\bibitem{ikh}D.J.Hinde, M.Dasgupta,  A.Mukherjee, Phys.Rev.Lett. {\bf 89}, 282701 (2002).
\bibitem{Quaal1}
A.Diaz-Torres, G.G.Adamian, N.V.Antonenko, W.Scheid,
Phys.Rev.C {\bf 64}, 024604 (2001); Nucl.Phys.A {\bf 679}, 410 (2001); A.K.~Nasirov {\it et al.},
Eur.Phys.J.A {\bf 34}, 325 (2007).
\bibitem{Toke}
P.Gippner {\it et al.}, Z.Phys.A {\bf 325}, 335 (1986);
G.N.Knyazheva {\it et al.},  Phys.Rev.C {\bf 75}, 064602 (2007);
M.G.Itkis {\it et al.}, Nucl.Phys.A  {\bf 787}, 150c (2007).

\end{thebibliography}
\end{document}